\documentclass[prd,aps,tightenlines,superscriptaddress,nofootinbib,eqsecnum,amsfonts,amsmath,epsfig]{revtex4} 
\pdfoutput=1
\input{epsf} 
\usepackage{thumbpdf}
\usepackage{graphicx}

\def\be{\begin{equation}}
\def\en{\end{equation}}
\def\bea{\begin{eqnarray}}
\def\ena{\end{eqnarray}}

\def\hn{\hat{n}}
\def\hk{\hat{k}}

\def\hu{\hat{u}}
\def\hv{\hat{v}}
\def\vp{\vec{p}}

\def\xn2{\frac{x_n}{2}}
\def\F{\mathcal{F}}
\def\M{\mathcal{M}}

\begin{document}

\title{Building a stochastic template bank for detecting massive black hole binaries.}
\author{Stanislav Babak}
\email[]{stba@aei.mpg.de}
\affiliation{Max-Planck-Institut fuer Gravitationsphysik, Albert-Einstein-Institut, Am Muchlenberg 1, D-14476 Golm bei
Potsdam, Germany}
\date{\today}
\begin{abstract}
Coalescence of two massive black holes is the strongest and most promising
source for LISA. In fact,  gravitational signal from the final inspiral and merger will be 
detectable throughout the Universe. 
 In this article we describe the first step in the two-step hierarchical search
for gravitational wave signal from the inspiraling massive BH binaries. It is based on
the routinely used in the ground base gravitational wave astronomy method of filtering the 
data through the bank of templates. However we use  a novel Monte-Carlo based (stochastic)
method to lay a grid in the parameter space, and we use the likelihood maximized analytically over
some parameters, known as $\F$-statistic, as a detection statistic. We build a coarse 
template bank to detect gravitational wave signals and to make preliminary parameter estimation.
The best candidates will be followed up using Metropolis-Hasting stochastic search to refine
the parameter estimation. We demonstrate the performance of the method by applying it
to the Mock LISA data challenge 1B (training data set).
\end{abstract}

\maketitle

\section{Introduction}

ÊCurrent astrophysical observations show abundant evidence that almost all galaxies
contain (super)massive black hole (MBH) in their nuclei. Already in 60-th it was proposed
\cite{Zel'dovich:1964} to explain the enormous luminosity of quasars by accretion of gas on
MBH.   The common believe is that a large fraction of galaxies must have been quasars in the
past. The study of the kinematics of stars and/or gas in the central galactic regions suggest
existence of the dark compact object with a mass $10^6-10^9 M_{\odot}$, which is most probably 
MBH \cite{Tremaine:2002js}. Only accretion mechanism cannot explain the growth in mass of all BHs
from the quasar to the present day, it suggests that they were formed by merging smaller mass 
BHs \cite{Hopkins:2005fb}, and, in fact, we do observe many merging galaxies. The most notorious is a direct VLBA observation suggesting existence of two MBHs with a total mass $10^8 M_{\odot}$ 
 and an orbital separation $\sim 7 pc$ in the center of radio galaxy 0402+379 \cite{Rodriguez:2006th}. 
 
 Last week of inspiral of MBH binary followed by the merger forming a single fast spinning excited BH
 is the most powerful source of gravitational waves (GW). In fact we will be able to detect those 
 events throughout the Universe with LISA. LISA \cite{LISA:1998}  is a proposed space borne 
 gravitational wave observatory which will be launched in $2018+$ and it will be sensitive to GW
 signals in the range $10^{-4}-0.1\; Hz$.  It consists of three identical spacecrafts forming an equilateral 
 triangle on heliocentric orbit. 
 
 In this paper we will concentrate on detecting only inspiralling part of the GW from the 
 coalescing MBH binary. This part is quite well modeled analytically using post-Newtonian (PN)
 approach, which is in fact iterative solution of the Einstein equations for two-body problem
 using expansion in small orbital velocity $v$\footnote{In fact it is expansion in $v/c$, but throughout
 the paper we work in geometrical units $G=c=1$.}. We restrict our attention here to the
 non-spinning BHs moving in the quasi-circular trajectory. Furthermore, we use restricted  
 PN waveform (keeping only leading order amplitude of GW signal) with the phase defined
 up to 2-nd PN order \cite{Blanchet:1996pi}.  It is not the most accurate today's waveform, but the 
 purpose of the current work is rather to describe data  analysis algorithm and to demonstrate
 its  performance, this work can be extended to more sophisticated models. Our conventions for the
  LISA model are the same as adopted for the mock LISA data challenge (MLDC) \cite{Arnaud:2006gm,
  Arnaud:2006gn}.

 This paper describes the first step in two-step hierarchical search method. It is based on constructing
 the coarse grid of templates and (matched) filtering the data through each template. The candidates
 will be taken as an input to the second stage of the search which utilizes simplified
 Metropolis-Hasting stochastic search. Before we go into the detailed description of our method we should mention the available alternatives. 
 
The most successful method so far is based entirely on the Metropolis-Hastings stochastic search
\cite{Cornish:2007jv,  Cornish:2006ms, Cornish:2006dt}. 
We cannot call it Markov chains as the method violates Markovian property (it uses previously available
information in some proposal distributions for the further jumps). The success of this method lies 
in a range of proposal distributions, those utilize symmetries in the detector's response function 
and global properties of the likelihood surface. This, in combination with simulated and frequency
annealing schemes, makes this method very fast (can be run on a laptop) and robust.  However in presence 
of several sources one has to hunt signals one by one: finding the signal, estimating its parameters,\
removing it and searching for the next one. This somewhat diminishes efficiency of this method.

Another method \cite{Brown:2007se} is closer to the proposed in this paper and it uses three stages: (i) construct time-frequency map of the data, identify the chirp and perform $\chi^2$ fitting to estimate masses and
coalescence time (ii) use the template grid (based on \cite{Babak:2006ty}) to do matched filtering search, and
(iii) use Metropolis-Hastings stochastic search as follow up on the candidates from the previous steps. 
This search method carries a lot of similarities with what we will describe below, but we should mention 
that our implementation is completely different. The advantage of using multi-stage
methods described above and presented in the paper is estimation of the number of signals present 
in the data. By mapping likelihood throughout the whole parameter space using template grid we get straight
away the set of all candidates (triggers, clustered in the time of coalescence, which crossed a preset 
signal-to-noise ratio (SNR) threshold).

Now we are ready to discuss our method. The main novelty of which is the new template bank: 
we have used a stochastic method to build it and this is the \emph{first} implementation. The 
focus of this paper will be on the construction of this template bank.
The main idea is trivial: you throw templates in the parameter space and compute
match (also called overlap) between trial template and templates in the bank, you add the trial template 
to the bank if the match is below a preset minimal match. Overlap is defined as a Hermitian inner product
between two normalized signals:

\begin{equation}
(\hat{s} | \hat{h}) \equiv 2 \int_0^{\infty} \frac{\tilde{s}^*(f)\tilde{h}(f) + 
\tilde{s}(f)\tilde{h}^*(f)}{S_n(f)}\; df\,.
\end{equation}
Here  $S_n(f)$ is the one-sided noise power spectral density, $\tilde{s}$ and $\tilde{h}$ are normalized signals
in the Fourier domain, and finally, the normalization means $(\hat{s} | \hat{s}) = (\hat{h} | \hat{h}) = 1$.

The template bank can be seen as a grid in the parameter space and the coarseness of the grid is
defined by the minimal match ($MM$) between nearby templates (proper distance between the neighboring 
points on the grid). For the ground based interferometers $MM$ is chosen as a compromise between computational power and the tolerated loss in detection. In the case of LISA, the signals are usually strong
and detection is not a problem, so our choice of $MM$ is governed by the speed of computation and by
the  closeness of the estimated parameters to the true values. Given $MM$, we want to construct 
template bank optimally: minimize number of templates while trying to keep overlap between neighboring
templates close to $MM$. The problem of template placement was attacked in gravitational wave 
community by several groups\footnote{Also private communication with C. Messenger and  I. Harry who also recently started to work on the stochastic banks, but use different implementation.} \cite{Babak:2006ty}, \cite{Prix:2007ks}. Previous banks
aimed at the (sub)optimal placement using metric induced on the parameter space. The main problem one 
encounters in those methods is the search for the coordinate frame which would be close to Cartesian if 
parameter manifold is flat (or almost flat) and which makes ambiguity contours\footnote{Ambiguity contours are contours of constant overlap in the parameter space} close to circles. Even if one
overcomes this problem, there is still a need to deal with boundaries which are usually overpopulated with
templates. In addition, attempt to cover parameter space completely (without holes) inevitably forces us
to place template with overlapping ambiguities contours, which leads to rather unnecessary  large number 
of templates. Instead, while using stochastic method, we will allow for small holes in  covering 
parameter space but increase slightly $MM$.

There could be several different ways to construct the stochastic bank in practice. We describe a 
particular way of building such a bank in the Section~\ref{Bank}, and this is the main result of this 
paper. Then we use the constructed bank to perform filtering of the noiseless data, the results 
are presented in the Section~\ref{Search}. Here we consider the simplified signal from MBH binary, which is restricted 2PN model adopted in MLDC \cite{Arnaud:2007jy} and summarized in the Appendix~\ref{appendixA}.
 The signal depends on five intrinsic parameters: masses ($m_1, m_2$), time of coalescence ($T_c$) and sky positions ($\theta_s, \phi_s$) and four extrinsic. 
In our search we have used $\F$-statistic \cite{Jaranowski:1998qm} to maximize over the extrinsic parameters: inclination ($\iota$), polarization ($\psi$), luminosity distance ($D_L$) and initial orbital phase ($\phi_0$), 
and we use iterative correlation method
to identify the time of coalescence. Effect of the coalescence time on the ambiguity contours is discussed in the 
Appendix~\ref{appendixB}. We summarize the key points of this paper in the concluding Section~\ref{summary}.

\section{Building stochastic bank}\label{Bank}

Let us repeat the main idea behind building the stochastic template bank. We start with choosing 
a random point in the parameter space and this is our first template in the bank. 
Then we loop over $N$ trials and for each trial (randomly chosen point in the parameter space within the prior) we 
compute overlap between the trial template and each template in the bank, if overlap is below $MM$
we add the point in the template bank, and reject it otherwise. If $MM$ is high, say above $0.95$
one can usually use metric induced on the parameter space to compute distance between trial and the 
neighboring templates in the bank:
\bea
ds^2 = g_{ij}\delta\lambda_i \delta\lambda_j.\label{}
\ena
Here $g_{ij}$ is the metric
\bea
g_{ij} = \left(\frac{d\hat{h}}{d\lambda_i} | \frac{d\hat{h}}{d\lambda_j}\right),
\ena
where we have assumed that the waveform $h(t, \lambda_i)$ depends on the  parameters
$\lambda_i$ and $\delta\lambda_i$ is mismatch in the parameters. For closely placed templates, overlap, $\mathcal{O}$ can be approximated as 
 \bea
 \mathcal{O} = 1 - ds^2,
\label{approxmetr}
 \ena 
 however if we increase the coarseness of the grid
 the quadratic approximation is not sufficient and using this approximation usually overestimates the 
 overlap, even more, the metric gives only local measure of distances between templates, whereas overlap is a global measure and it is important for placing templates, as we will show later in this section. For our search 
 we have chosen minimal match to be $0.88$ and it is too low to use metric (local quadratic approximation) in
 estimating overlap. 
 
 Computing the overlap directly is computationally expensive procedure. To tackle this problem we use two 
tricks. Firstly, we have assumed that the coalescence time is 5 hours after the observation period which 
is taken to be 1 year. The duration of observation is motivated by the full cycle in LISA motion and therefore 
in the Doppler modulation of the signal. The choice of the coalescence time was mainly
governed by our desire to have the highest frequency of templates below 1 mHz, which allows us to use 
long wavelength approximation to the LISA's response. We have used only $X$ TDI channel, unequal arm
Michelson combination (see Apendix~\ref{appendixA} for more details on the signal's model). 
Restricting the highest frequency of the signal also enabled us to use low sampling 
rate. Finally, we manage to compute about 200 overlaps per second on a single core of Intel DUO 2.
The second trick is to estimate the density of templates in the parameter space, then split it in the 
regions of approximately equal and small number of templates, and to populate each sub-region 
separately using cluster of computers. Small number of templates allows us to do many trials and
have a reasonable overall coverage. Having about equal expected number of templates in each sub-region
balances the computational load in parallel computing.

We have noticed that for a fixed masses the ambiguity contours which correspond to $\mathcal{O} = 0.88$
do not change significantly for different sky locations. This implies that we can use a fix sky position, lay
template grid in $m_1$ -- $m_2$ space and it will be a valid grid (with good approximation) for any 
sky position.  Moreover  $MM=0.88$ is too low for placing templates in the sky position, ambiguity 
contours are very big, and we use instead a uniform grid on the sphere with approximately 2700 points. 
The separation of points on the sphere was taken as double expected error in the determination of the sky location as defined by the inverse Fisher matrix. We have also found that other (extrinsic) parameters do 
not affect the ambiguity contours,  meaning that contours stayed almost the same over several arbitrary 
chosen values, so we use a particular choice: ($\iota = 0,\; \psi=0,\; \phi_0 =0$). All in all, we need to place templates in $m_1$ -- $m_2$ plane 
only. To search for coalescence time we slide templates across the data and look for the maximum of correlation. 
We have found that ambiguity contours are rather sensitive to the particular choice (fixed) of coalescence 
time, they change not only in size but they also rotate. The chosen coalescence time is not the most conservative, 
but, as it follows from the Section~\ref{Search}, this is cured by iterative maximization of $\F$-statistic 
over this parameter during the search. One can find more details on this in the next section and in the Appendix~\ref{appendixB}.

\begin{figure}[ht]\label{ambig}
\includegraphics[height=0.35\textheight, keepaspectratio=true]{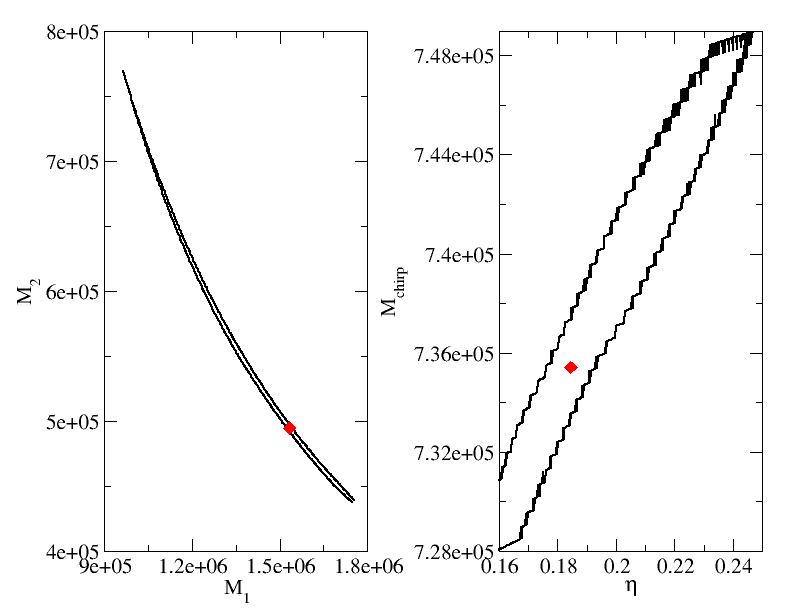}
\caption{Global ambiguity contours: left panel in $m_1 - m_2$ and right panel in $\eta - \M$ coordinates.
The central point is depicted as a red diamond.}
\label{Ambig1}
\end{figure}

As mentioned above, first we need to estimate the density of templates in $m_1$ -- $m_2$. We have found 
that $m_1, m_2$ are rather not convenient coordinates. In the Figure~\ref{Ambig1} we have shown the 
typical \emph{global} ambiguity contours of match $0.88$ (we will explain meaning of ``global'' a bit later) 
in $m_1 - m_2$ space and in $\eta  - \M $ space, here masses are red-shifted masses in units of solar mass, 
$\eta = m_1m_2/M^2$ -- symmetric mass ratio, $\M= M \eta^{3/5}$ is a chirp mass and $M = m_1+m_2$. 
One can see that the ambiguity contour in masses is almost one-dimentional structure, and indeed in the 
$\M - \eta$ space the chirp mass changes by less than $3\%$ over the whole range of $\eta$. We should mention that we restrict the parameter space to the prior adopted in MLDC: $m_1 \in [10^6, 5\times 10^6]
M_{\odot},\; m_1/m_2 \in [1,4]$. The size of the contours does change across the parameter space, in addition
they also rotate, but the angle of inclination to $\eta$-axis is always small. This makes perfect sense, because
the first, dominant, term in the gravitational wave phase depends only on $\M$. I think those arguments are convincing enough
to adopt $\eta, \M$ as the coordinate frame for estimating density of templates.

\begin{figure}[ht]\label{ambig}
\includegraphics[height=0.3\textheight, keepaspectratio=true]{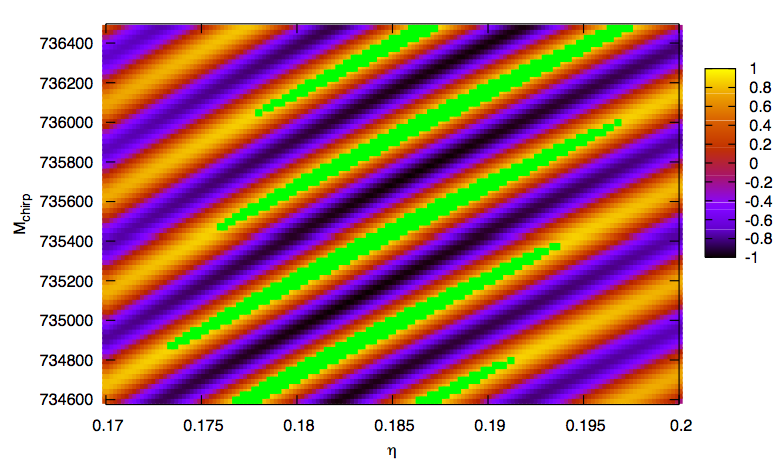}
\caption{Internal structure inside the global ambiguity contour. It is 3-d plot projected onto
$\eta - \M$ plane and z-axis (overlap with central point) is color-coded. Overplotted green point
are those with overlap above $MM=0.88$.}
\label{MiniAmbig}
\end{figure}

Before we go further we need to explain the meaning of ``global'' ambiguity contours. Inside those global
contours overlap exhibits damping oscillations, which maximum amplitude reaches $1$ at the central
point and damps to $MM = 0.88$ at the boundaries of the global contour. In the Figure~\ref{MiniAmbig}
we zoom into a part of the global contour around the central point ($\eta = 0.1845,\; \M= 735423$).
The green points correspond to the overlap above $MM$, and the overlap with central point is color-coded.
One can see the separated maxima within the global contour. 
This is not something new, such structure was first described in
\cite{Balasubramanian:1995bm}, and later in context of LISA in \cite{Cornish:2006ms}. Let us emphasize an
important point, the ambiguity ellipse described locally by the metric on the parameter space would correspond
only to the green ellipse around the central point in the Figure~\ref{MiniAmbig}, and ignore other maxima.
Taking only the central small ellipse as ``area'' of the template in the parameter space will seriously
overpopulate the parameter space with templates. The size, orientation and hight (maximum overlap) of each
small ellipse within the global contour as well as their separation depend on $m_1, m_2, T_c$. 
We will present more detailed study of the global properties of the likelihood using $\F$-statistic (rather than 
overlap) in the separate publication.

\begin{figure}[ht]\label{ambig}
\includegraphics[height=0.3\textheight, keepaspectratio=true]{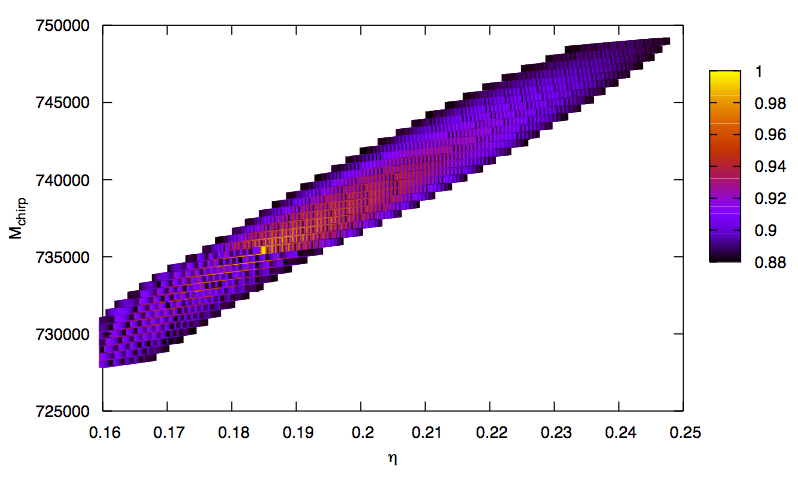}
\caption{Distribution of the overlap inside the global ambiguity contour. It is 3-d plot projected onto
$\eta - \M$ plane and z-axis (overlap with central point) is color-coded.}
\label{Ambig2}
\end{figure}

The distribution of the overlap inside the global contour is presented in the Figure~\ref{Ambig2}. 
Note that orientation of the global ambiguity contour is different from the orientation of the local ellipses inside
the global contour.

\begin{figure}[ht]\label{ambig}
\resizebox*{.45\textwidth}{.4\textwidth}{\includegraphics[keepaspectratio=true]{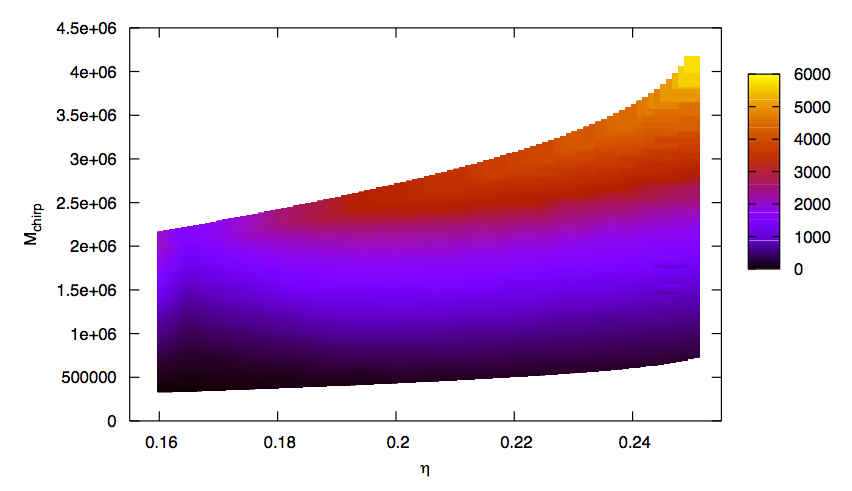}}
\resizebox*{.45\textwidth}{.4\textwidth}{\includegraphics[keepaspectratio=true]{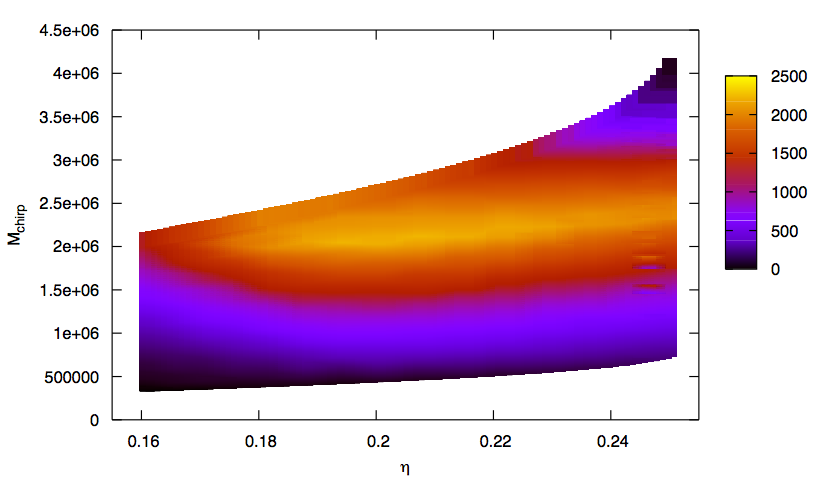}}
\caption{Interpolated area of template. Left panel: area of template as defined by global ambiguity 
contour across parameter space; right panel: useful area of template (part of the global ambiguity
contour which lies inside parameter range defined by priors).}
\label{Area}
\end{figure}

Return back to the estimation of the template density in the parameter space. First of all 
we will define the area of a template as an area inside the global ambiguity contour. This is clearly 
overestimation of the area of a template, since there are ``holes'' inside the global structure,
but after all we want to split parameter space in the regions of an approximately equal number
of templates, and not to do rigorous template counting. We compute area of 336 uniformly distributed 
templates. Then we compute number of templates as in \cite{Owen:1998dk}, assuming that 
$\sqrt{||\rm{det}\;g_{ij}||} \sim (1-MM)/A_{t}$, where $A_{t}$ is interpolated (coordinate dependent) 
area of a template:
\bea
N \sim \frac{\int \sqrt{||\rm{det}\;g_{ij}||} dx dy }{(1 - MM)} \approx \int \frac{dx\; dy}
{A_t} \label{N}
\ena
Again, the absolute number is not very important for us here, what important is how it changes
(accumulative number) across parameter space. In the Figure~\ref{Area} we plot the interpolated 
total area of templates (in the left panel) and the area of templates which falls inside the parameter 
region defined by priors (in the right panel).
One can see that (i) contours of equal area again correspond to approximately levels
of constant $\M$; 
(ii) the splitting of the parameter space
based on the left or right panel is not the same. We have used the left panel (total area of the 
template) to do splitting of the parameter space, because we used assumption of the elliptical
shape of the area in (\ref{N}) which is better approximation for the total area. However we will 
keep in mind that expected deviation in number of templates from the mean will be higher in 
$\{0.16, 5.e5\}$ and in $\{0.25, 4.e6\}$ corners of the $\eta - \M$ parameter space.

\begin{figure}[ht]\label{ambig}
\includegraphics[height=0.3\textheight, keepaspectratio=true]{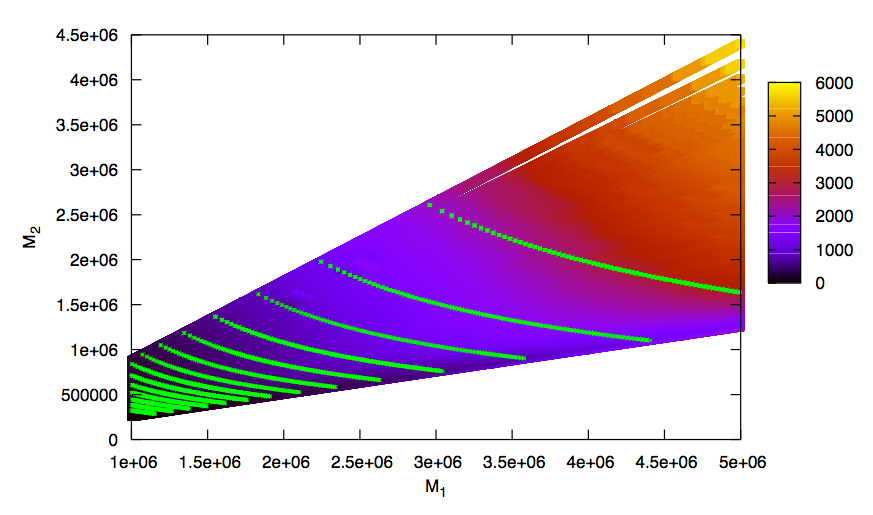}
\caption{Splitting of the parameter space $m_1 - m_2$ in 15 sub-regions. Green crosses correspond
to boundaries, color-coded z-axis is area of a template.}
\label{split}
\end{figure}

By looking at the cumulative number of template we have split parameter space in 15 sub-regions, those
correspond to contours of equal $\M$. In the Figure~\ref{split} we show this splitting 
in $m_1 - m_2$ space.

\begin{figure}[h]\label{ambig}
\resizebox*{.45\textwidth}{.35\textwidth}{\includegraphics[keepaspectratio=true]{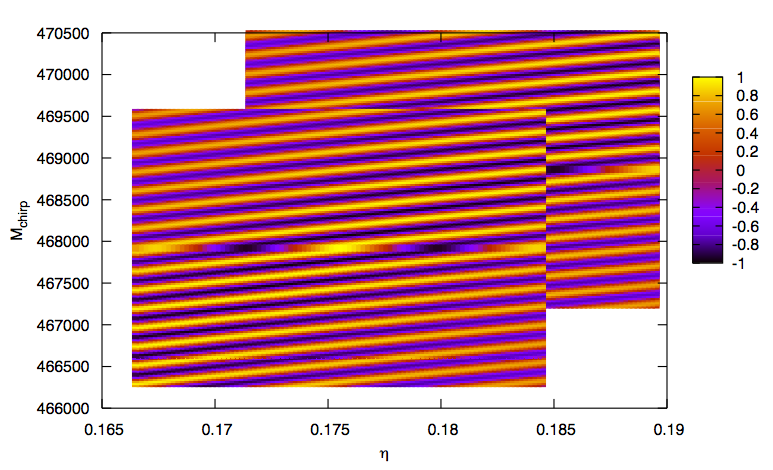}}
\resizebox*{.45\textwidth}{.35\textwidth}{\includegraphics[keepaspectratio=true]{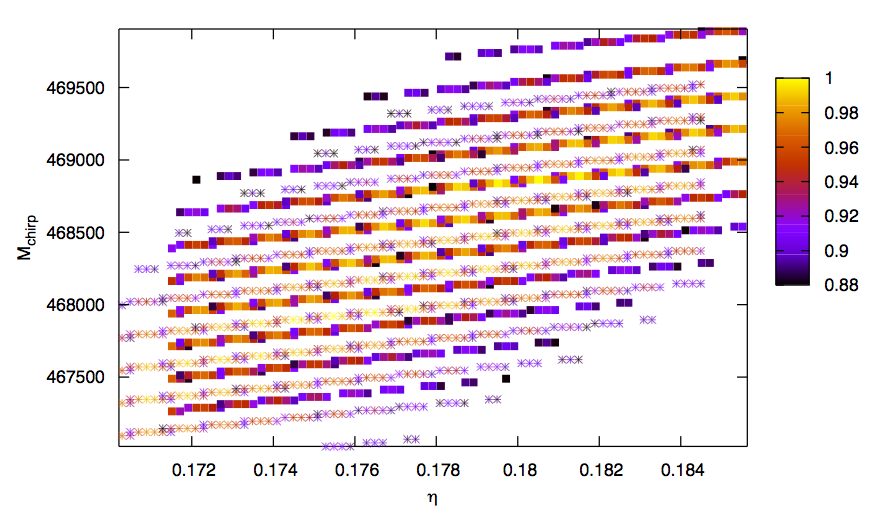}}
\caption{Internal structure of two overlapping global ambiguity contours corresponding to two different templates. Left panel: overlap structure around the central points, the yellow maxima of one template
fall onto the local minima (dark blue) of other. Right panel: the same as the left panel but here we keep 
only points with the overlap exceeding $MM=0.88$, crosses and boxes corresponds to the ambiguity ellipses
of two different templates ($\{0.175, 467921\},\; \{0.18, 468864\}$) .}
\label{placement}
\end{figure}

We have populated each sub-region separately in parallel, we terminated Monte Carlo when number of 
trials exceeded square of the current number of template in the bank. The final number of templates in each 
sub-region is $[495,\; 346,\; 352,\; 375,\; 404,\; 408,\; 416,\; 432,\; 462,\; 479,\; 489,\; 517,\; 517,\; 
532,\; 510]$, with mean value $449$ and standard 
 deviation $63$. The number of templates is approximately 18 times more than predicted by a global 
ambiguity contours, and as mentioned earlier this is due to existence of the hollow spaces inside
the global ambiguity contours. In the Figure~\ref{placement} below we show one of such cases. There the
global contours overlap but the local maxima of one template fall onto local minima of other template.

By filling up each of the sub-regions separately, we over-populate the boundaries between sub-regions
(they were populated from both sides).
It is not too much, $\sim 10\%$, for all 14 boundaries. One can get rid of some templates by computing overlaps
between templates in near-boundary areas, however this procedure might crate ``holes'': the removed template,
say $a$, which has overlap with $b$ above $MM$, might have covered part of the space which is not
covered by $b$. Therefore one needs to do an additional Monte Carlo to cover the newly created 
``holes'' in the  near-boundary regions.  

To conclude this Section let us mention that if efficiency of the method is not a concern, then one can
try to populate the whole parameter space without splitting or by making an arbitrary splitting.
Alternative would be to populate with higher density  regions of the parameter space which would 
more likely contain the signal from the astrophysical point of view. 
However studying shapes, size and orientation of the global ambiguity contours and local maxima
is very useful exercise which helps to conduct the search and to understand the results.

\section{Application of the stochastic bank to MLDC.}\label{Search}

We have used noiseless training data set 1B2.2 to filter through the template bank for the testing purposes
 and to avoid the bias in parameters estimation caused by presence of the noise. Even though we have
used the training data set (parameters of the signal are known), we have conducted the blind search
(pretending we do not know signal's parameters). The prior knowledge was that there is only one signal
with $T_c = 400 \pm 40$ days and SNR between 20 and 100 in a single TDI channel. For the search 
we have used so called rigid adiabatic approximation to the LISA's response (described in the 
Appendix~\ref{appendixA}) and combined $\F$-statistic for two channels with uncorrelated noise, 
$X$ and $Y-Z$  \cite{Krolak:2004xp}. 

We should explain here the search for coalescence time, it is similar to one used in \cite{Cornish:2006ms}. 
We start with assuming some $T_c$, usually at the beginning of the prior, and instead of computing 
inner products we compute correlation for different time lags:
\bea
C_{h,s}(\tau) = 2 \int \frac{\tilde{s}(f) \tilde{h}^*(f)}{S_n(f)}e^{i 2\pi f \tau} df
\ena
Inner product is just a correlation at zero lag. Then $\F$-statistic becomes an array, 
$\F(t_k = k \Delta t)$, and we search for its maximum value. By doing this we are trying to fit the 
signal's intrinsic phase, assuming the
wrong amplitude modulation (antenna beam function depends on the choice of $t_0$ and $T_c$). Then we use
a new value of $T_c$ (lag at which was maximum of $\F(t_k)$) and repeat the procedure, usually it converges 
to maximum of $\F(t_k)$ at zero lag within few iterations. This procedure works very well and makes 
search for $T_c$ very efficient, but due to the coarseness of data sampling the best estimation of $T_c$
is a nearest multiple of $\Delta t$, for the current search we have used $\Delta t = 120$ sec.  

\begin{figure}[ht]\label{ambig}
\includegraphics[height=0.35\textheight, keepaspectratio=true]{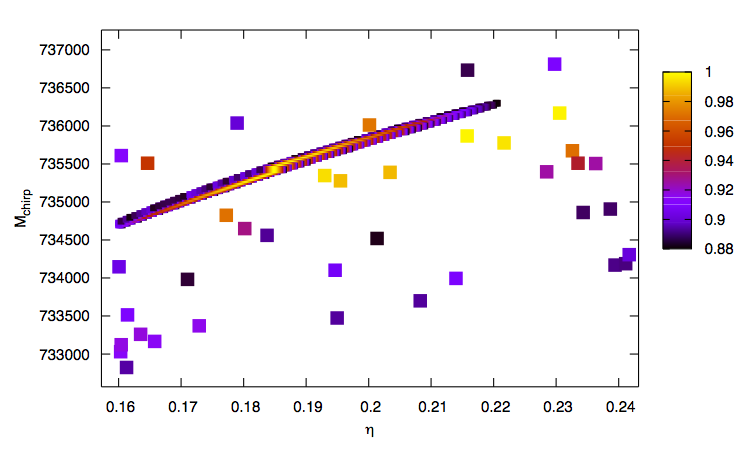}
\caption{Results of the search: we overplotted an ambiguity contour at the true point (the overlap value is color-coded)
with points of the bank (large boxes) which gave high values of $\F$-statistic. For those points color-coding 
denotes ratio of found $\F$ to $\F_{true}$, value of $\F$-statistic for the template with true parameters.}
\label{searchAmb}
\end{figure}

 Previous studies shown that the coalescence time correlates with the chirp mass, this, usually undesired feature,
rather helps us to improve efficiency of the bank. As we have mentioned above and discuss in the 
Appendix~\ref{appendixB} ambiguity contours and local maxima depend rather strongly on choice of $T_c$
for a given $MM$, however allowing maximization over $T_c$ (and due to correlation  between $T_c$ and $\M$)
different points from/besides the expected give a very high value of $\F$-statistic. We have tried to illustrate 
this in the Figure~\ref{searchAmb}. In there we have plotted the ambiguity contour (88\%)\footnote{Note that in this case local secondary maxima are lower than 0.88, so the local and global contours are the same} around the 
true point and points of the bank which gave high value of $\F$-statistic. Note that those points lie outside
the ambiguity ellipse, at the same time, as we will show soon, those points are off in $T_c$. We can use 
the correlation $\M - T_c$ in the second step of search to find the true parameter of the signal. 

So we have searched for four parameters: masses and sky location (with maximization over coalescence time).
The best  seven points are presented in the Table~\ref{sres}. The candidates are separated in two groups:
the first group corresponds to the true sky location of the signal and the second group shows the antipodal
sky position. It is known (see for example \cite{Cornish:2006ms}) that for the low frequency sources 
($L\omega_{gw} \ll  1$, $L$ is armlength) LISA's response function is quite symmetric with respect to the 
direct and antipodal sky positions. We have found quite well the sky location and relatively well the chirp mass.
Note that those secondary maxima have $\F$-statistic value very close to the maximal. So overall our bank did 
quite well. The candidates now can be passed to the second step of search.

Let us say few words about the second step of search. The second step is similar to described in 
\cite{Cornish:2006ms}, but we do not need anymore frequency annealing. Metropolis-Hastings stochastic search
can be simplified since we are already close to the true point. We use simulated annealing as it helps us
to move off the local maxima. Note that we know a global structure of likelihood (it is the same as 
internal structure of global ambiguity contours) in that region of space from the bank building stage, so we can 
use this information to construct an efficient proposal distribution (similar to "island jumping" described in
\cite{Cornish:2006ms}). So in principle we need here only three proposal distributions: (1) jumps along
eigen directions of  Fisher matrix (to search in vicinity of the local maxima), (2) jump between local maxima (3) check direct
and antipodal sky locations. We have successfully\footnote{Actually "semi-successfully" we have made an error
in estimating extrinsic parameters, and we have used wrong value of siderial year. In addition, due to lack of time,
the chains in the second step of search were not long enough.} implemented this search in the second round of MLDC 
and results are summarized here \cite{2007arXiv0711.2667B}. 

\begin{table}
\caption{\label{sres}
True parameters and seven best points of the search. Two groups represent correct and antipodal 
sky locations. The columns are: primary mass in $M_{\odot}$, secondary mass in $M_{\odot}$, symmetric mass 
ratio, chirp mass in $M_{\odot}$, ecliptic longitude in radian, ecliptic latitude in radian, time of coalescence in 
seconds, value of $\F$-statistic}
\begin{ruledtabular}
\begin{tabular}{ccccccccc}
name & $m_1$ & $m_2$ & $\eta$ & $\M$
 & long & lat & $T_c$ & $\F$ \\
\hline
True & 1532832  &  494760  & 0.1845 & 735423 & 5.71 & 0.158 & 34869697.9  &  3344.1 \\
\hline
Candidate & 1265090  &  581679  &  0.216 &  735869  &   5.70  &  0.20  &  34880160.0 & 3338.4 \\
Candidate & 1135330  &  640146  &  0.231 &  736169  &   5.70 &   0.14  &  34882080.0  & 3337.7 \\
Candidate & 1397630  &  534699 &  0.200 &  736012  &  5.78 &  0.03 &  34868520.0  & 3253.2 \\
\hline
Candidate & 1214270   &   602610 &  0.222  &  735776 &  2.60 &  -0.14 &  34882800.0  & 3329.4 \\
Candidate & 1457870  &  515425  &  0.193  &   735347 &  2.60 &  -0.14 &  34874400.0  & 3325.6 \\
Candidate & 1435940  &   521841 &  0.195 &   735278  &  2.51 &  -0.14 &   34877160.0  &  3306.0 \\
Candidate & 1368550  &   543364 &   0.203 &  735389 &  2.51 &   -0.14 &   34879800.0  &  3303.1\\
\end{tabular}
\end{ruledtabular}
\end{table}

\section{Conclusion}\label{summary}

In this paper we have presented a new (stochastic) method to construct two-dimensional 
template bank for detecting gravitation wave signal from the inspiralling non-spinning MBH binaries.
The main idea is to throw points randomly onto the parameter space and keep only those
with overlap below a preset minimal match. We have shown how to implement such a method efficiently
by splitting parameter space into sub-regions which can be covered with approximately the same number 
of templates. As a bonus we have also learned about local and global properties of the likelihood (or overlap)
across the parameter space which is useful in designing search algorithms. Even though we have used this method 
in two-dimensional space, it can be extended to higher dimensional parameter manifolds. 
We used this bank to filter noiseless signal and have shown that the bank is efficient to capture the signal
and to make preliminary parameter estimation. Candidates together with information about the properties of 
the likelihood in the vicinity of those points should be used in the second (follow up) step of the search.

We should mention that while studying the local and global properties of the overlap function, we 
have realized the importance of such study, and have started investigating global properties 
of the likelihood surface for the signals from Galactic binaries and inspiralling MBH binaries. There we rather
use $\F$-statistic instead of overlap. The results of this research will be published in the forthcoming 
paper.

\section{Acknowledgment}
Author would like to thank Bruce Allen, who is behind the idea of placing templates stochastically;
and also special thanks to Ed Porter for the useful discussions. This work was supported by DLR (Deutsches Zentrum f\"ur Luft-  und Raumfahrt).

\appendix

\section{Waveform}\label{appendixA}

Here we give a brief summary of the waveform we have used in the search and for constructing 
the template bank. 
We start with the waveform in the radiative frame, following \cite{Finn:1992xs}:

\bea
h_{+}^{rad} = \frac{2M\eta}{D} (M\omega)^{2/3} h_{+}^o \cos2(\Phi + \Phi_0)\\
h_{\times}^{rad} = \frac{2M\eta}{D} (M\omega)^{2/3} h_{\times}^o \sin2(\Phi + \Phi_0),
\ena
here $\omega$ is the instantaneous orbital frequency and $\Phi$ is the orbital phase.
Note that we are using restricted PN waveform. The two polarization amplitudes are
\bea
h_{+}^o &=& 1 + \cos^2i\\
h_{\times}^o &=& -2\cos i
\ena
and the inclination angle $i$ is between the orbital angular momentum and the direction to the source
in Barycentric frame.
It is convenient to use a complex representation of the waveform:
\bea
h_{+} &=& \frac{M\eta}{D} (M\omega)^{2/3} h_{+}^o e^{2(\Phi + \Phi_0)} + c.c.\\
h_{\times} &=& \frac{M\eta}{D} (M\omega)^{2/3} (-ih_{\times}^o) e^{2(\Phi + \Phi_0)} + c.c.
\ena
We use convention for Fourier transformation which agrees with ``fftw'':
\[
\tilde{h}(f) = \int h(t) e^{-2\pi ift}\;dt
\]
In the solar system barycentric frame (SSB) the waveform is
\bea
h_{+} &=& h_{+}^{rad}\cos 2\psi + h_{\times}^{rad}\sin 2\psi\\
h_{\times} &=& - h_{+}^{rad}\sin 2\psi + h_{\times}^{rad}\cos 2\psi
\ena
We need to compute TDI streams. The single arm response is \cite{1999ApJ...527..814A}
\bea
y_{slr} = (1 + \hk\hn_l)\left[ \Psi_l(t-L-\hk\vp_s) - \Psi_l(t-\hk\vp_r)  \right],
\ena 
where
\bea
\Psi_l = \frac1{2} \frac{(\hn_l)_i h^{ij} (\hn_l)_j}{1 - (\hk\hn_l)^2}.
\ena
Here $L$ is the armlength (assumed to be constant), $\hn_l$ is a unit vector along the interferometer's arm, $\hk = - \hn$ is the direction of the
wave propagation, and $\vec{p}_j = \vec{R} + \vec{q}_j$ is position of $j$-th spacecraft in (SSB), 
 $\vec{R}$ is position of the guiding center and  $\vec{q}_j$ is the spacecraft's position with respect to the 
 guiding center. The field of gravitational wave is 
\bea
h^{ij} &=& h_{+} \mathcal{E}^{ij}_{+} + h_{\times}\mathcal{E}_{\times}^{ij}\\
\mathcal{E}^{ij}_{+} &=& u^i u^j - v^i v^j,\;\;\; 
\mathcal{E}^{ij}_{\times} = u^iv^j + u^j v^i
\ena
Here $\hu,\; \hv$ define the polarization basis in the SSB:
\bea
-\hk &=& \hn = \{ \sin\theta_S \cos\phi_S, \sin\theta_S\sin\phi_S, \cos\theta_S \}\\ 
\hu &=& \{\cos\theta_S\cos\phi_S, \cos\theta_S\sin\phi_S, -\sin\theta_S\}\\
\hv &=& \{ \sin\phi_S, -\cos\phi_S, 0\}.
\ena
Taking the above into account we get
\bea
\frac1{2} (\hn_l)_i h^{ij} (\hn_l)_j &=& U_l h_{+}  + V_l h_{\times} \\
&=& \frac{M\eta}{D} (M\omega)^{2/3} e^{2(\Phi + \Phi_0)} 
\left[ U_l F_u + V_l F_v\right] + c.c.
\ena
and
\bea
U_l &=& \frac1{2} \left[ (\hn_l \hu)^2 - (\hn_l\hv)^2\right],\;\;\;
V_l = (\hn_l\hu)(\hn_l\hv)\\
F_u &=& h_{+}^o\cos 2\psi - ih_{\times}^o\sin 2\psi,\;\;\;
F_v = - h_{+}^o\sin 2\psi - ih_{\times}^o\cos 2\psi
\ena

Then one can compute the response for a particular TDI combination
using the phase expansion around the SSB time which leads to the following
expression for  the single link:
\bea 
y_{slr} = i \frac{M\eta}{D} (M\omega)^{2/3}\left[ U_l F_u + V_l F_v \right]
e^{i2[\Phi(t) + \Phi_0 - \omega \hk\vec{R}]}
(\omega L) sinc\left[\omega L(1-\hk\hn_l)\right]
e^{-i\omega (L + \hk(\vp_s +\vp_r))} + c.c.
\ena
This approximation we call rigid adiabatic approximation.
In particular for the $X$-channel (unequal arm Michelson) we have
\bea
X^{gw} = \frac{M\eta}{D} (M\omega)^{2/3}e^{i2[\Phi(t) + \Phi_0 - \omega \hk\vec{R}]}
\sum_{j=2,3}  \left[ U_j F_u + V_j F_v \right] \chi_j^X
\ena
actually, only $\chi_j$ is different for different TDIs. For $X$ we will write it explicitly:

\bea
\chi_2^X &=&  -2 x \sin{x}\left\{ sinc\left[ \frac{x}{2}(1-\hk\hn_1)\right] e^{-ix} + 
		sinc\left[ \frac{x}{2}(1+\hk\hn_1)\right]\right\} e^{-i\frac{x}{2}[3 + \hk(\vec{q}_0 + \vec{q}_2)]}
		\nonumber \\
\chi_3^X &=&  2 x \sin{x}\left\{ sinc\left[ \frac{x}{2}(1-\hk\hn_2)\right]  + 
		sinc\left[ \frac{x}{2}(1+\hk\hn_2)\right]e^{-ix}\right\} e^{-i\frac{x}{2}[3 + \hk(\vec{q}_1 + \vec{q}_0)]}.
\label{rad}
\ena
In the long wavelength limit $x= L\omega_{gw} \ll 1$, those become 
$$
\chi_2^X = -4 x^2,\;\;\; \chi_3^X = 4 x^2.
$$ 
We have used the long wavelength expression for constructing the bank, and we have used expressions
(\ref{rad}) for computing $\F$-statistic.

 \section{Effect of coalescence time on the ambiguity contours}\label{appendixB}
 
\begin{figure}[h]\label{ambig}
\includegraphics[height=0.35\textheight, keepaspectratio=true]{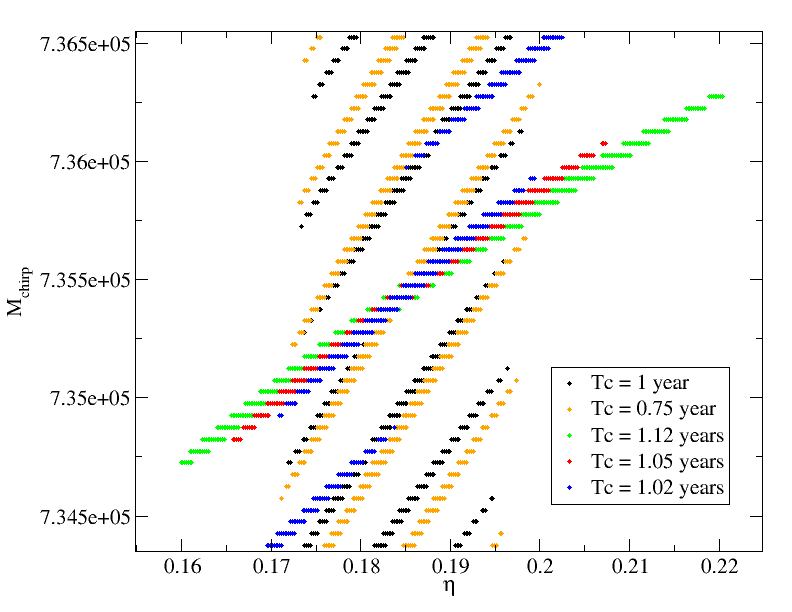}
\caption{Ambiguity ellipses for the different coalescence times.}
\label{ElTc}
\end{figure}

We have computed ambiguity contours around the true point of our search for the different 
fixed times of coalescence. In other words, we kept the sky location and extrinsic 
parameters constant and plotted ambiguity contour in $\eta - \M$ plane for the different
values of $T_c$. Results are presented in the Figure~\ref{ElTc}. One can  
notice several important points: (i) ellipses of the local maxima increase in size as we go to
larger $T_c$, (ii) ellipses of the local maxima rotate and distance between them changes,
(iii) the maximum overlap for the secondary maxima decreases as we increase $T_c$,
(iv) this is may be not obvious from this figure, but the global contour also slightly rotates.

All these points can be easily explained: by changing $T_c$ and keeping fixed duration of the observation
 we effectively change the frequency range of the signal: by moving $T_c$ further away we are observing
 the low frequency part of the inspiral only.  
 We know that the accuracy in the  parameter estimation drops
 for a low frequency source, therefore we see increase in the size of the local maxima for larger $T_c$. 
 Low frequencies also 
 correspond to the less relativistic regime of inspiral, and the phase of the wave is governed 
 largely by the leading order term
 which depends on $\M$ only, therefore, for large $T_c$, uncertainty in $\eta$ increases and ellipses are more 
 inclined toward the $\eta$-axis.  Once we start to see the end of inspiral (which is $\ge 1$ mHz),
 the majority of SNR comes from the last ten-to-hundred cycles (signal reaches most sensitive
 bandwidth of LISA, and intrinsic amplitude of the signal increases rapidly). Similar for overlap, the main
 contribution comes from the end of inspiral. If we observe the whole inspiral, the last $\sim$hundred 
 cycles can be fit by many signals with different parameters, therefore we see the set of secondary 
 maxima with the slowly damping strength (maximal overlap within the secondary ellipses) 
 
 This figure also suggest that for $MM=0.88$ the conservative choice (smallest ambiguity contour) corresponds
 to $T_c \approx 1.05$ years. But this would produce significantly larger number of templates, which 
 we believe is not necessary. This is because, as we discuss in the main body of the article, due to the
 maximization over $T_c$ and correlation of $T_c$ with $\M$. What is more important is to take into account
 this correlation on the second step of the search to identify the true parameters (primary maximum).

\bibliographystyle{apsrev-nourl}

\bibliography{refr}

\end{document}